\documentclass{ws-ijmpa3}
\def\be{\begin{equation}}
\def\ee{\end{equation}}

\begin{document}

\title{INTRODUCTION TO PINHOLE ASTRONOMY}

\author{Costantino Sigismondi}

\address{Department of Physics, University of Rome "La Sapienza" and ICRA, 
International Center for Relativistic Astrophysics, P.le A.Moro 2 00185 
Rome Italy}

\maketitle

\begin{abstract}
The observation of the Sun with a pinhole in order to check the first and the second Kepler's law has been realized for high school students. The settling of the experiment in order to minimize the errors of measure has permitted to verify the Rayleigh formula for obtaining sharpest images given the diameter of the pinhole.
\end{abstract}

\section{Introduction} 
Aristoteles already observed that far from an hole of whatever shape the image of the Sun above a screen was always a circle. For him it remained an unsolved problem. Leonardo da Vinci also worked on this topic, exploiting the dark camera principle mainly for drawing. Kepler in 1600 invented the term {\em dark camera} and used it for studying the partial phases of the total eclipse of the sun in Graz.\cite{lom}
Pinhole photography is nowadays a fine hobby, with several fans, and a conspicous literature about it was produced up to now.
The formula of Rayleigh describes the relationship between the pinhole diameter and optimum focal distance,\footnote{http://home.sol.no/~gjon/pinhole.htm} being $\lambda=5.5 \cdot 10^{-7}m$ the wavelength of the light:

\be
d=1.9\sqrt{\lambda f}.
\ee

We have found empirically this optimum value following a logical criterium, very instructive to be repeated.
\section{Measuring the eccentricity of Earth orbit}
The ellipticity of Earth orbit is the cause of the periodical changing of the apparent dimensions of the un. On January $4^{th}$ the diameter of the sun is 1952 arcsec with the Earth at perihelion, while on July $4^{th}$ it is 1888 arcsec at aphelion.
Being the apparent diameter of the sun inversely proportional to the Earth-sun distance we have the following equation for the eccentricity (first order approximated)
\be
e=\sqrt{1-\frac{d_{min}}{d_{max}}}.
\ee
From the above expected values we obtain $e=0.0167 \approx \frac{1}{60}$, but it is necessary to be able to measure the apparent diameter of the Sun with a precision well better than $3\%$ in order to obtain the correct value for $e$.
Moreover sampling montly the diameter of the Sun with such an accuracy it is possible to envisage the second Kepler law having a direct indication of the evolution of the magnitude of the radius vector with time.

\section{Improving the experimental conditions}
Starting with direct projection of the image of the Sun above a screen, it is immediate that such geometry is unfair because of the difficulty to settle the pinhole, and for measuring the focal distance. Note that we define as focal distance the distance between the pinhole and the screen (from centre to centre). Furthermore we use perfectly circularly shaped pinholes, having noticed before the blurring of the image with non circular pinholes.
The more simple geometry for reaching easily large focal distances is obtained using a mirror for projecting the Sun towards the pinhole, with mirror, screen and pinhole all groundbased.
In this way we could measure the diameter of the Sun with focal lengths up to 17 meters with a pinhole of only 7 mm of diameter, although the image appears weakly contrasted upon a white screen under daylight. 

\section{Settling the experimental errors}

At a focal distance of 17 m the daily apparent motion of the Sun (15 arcsec per second) is a strong source of error of measure.
In fact the diameter of the Sun is $17~{\rm m}\times \tan{\rm( 30 ')} \sim 16$ cm and its motion upon the screen is $17~{\rm m}\times \tan{\rm( 15 '')} \sim 1.3~~$mm. 
Another source of error is the contrast of the image on the screen. The {\sl dark camera} should be the better solution, and the projection of the image into closed rooms gives very good results.
Due to the self-absorption of the atmosphere of the sun, the limb of the Sun is darker than its centre. It appears clearly even observing the Sun with a 5 mm pinhole with not sharped (on optimum focus) boundaries of the image.
In addition to the darker limb there are quantum effects due to the diffraction of the light; a small quantity of radiation coming from those limbs goes outside its geometrical limit by an angular amount of 
\be
\theta {\rm [arcsec]} =\frac{120}{d{\rm (mm)}}.
\ee

\section{The systematic error of the pinhole apparatus}
The next step is to compare the measure of the ratio between the apparent diameter of the Sun and the focal distance.
It is equal to the ratio between the diameter of the Sun and the Earth-Sun distance. In fact the angles subtended by the Sun and its image with respect to the centre of the pinhole are opposite from the vertex (the pinhole).
We observed that some of the measures were greater than the expected values and also that the measures with smaller focal length suffered of a greater eccess with respect to the expected values.
We deduced that the responsible was the finite opening of the pinhole.
In a basic model we can draw three parallel cones starting respectively from two opposite edges of the pinhole and from its center, as in figure 1. Far from the pinhole the cones start to intersect each other. Putting a screen perpendicularly to the axes of the cones, the regions of overlapping of the three cones are expected to appear brighter. The remaining weaker region is, by construction, a circular corona of radius equal to the pinhole radius $r$, that we call {\em confusion circle}.
\begin{figure}
\centerline{\psfig{file=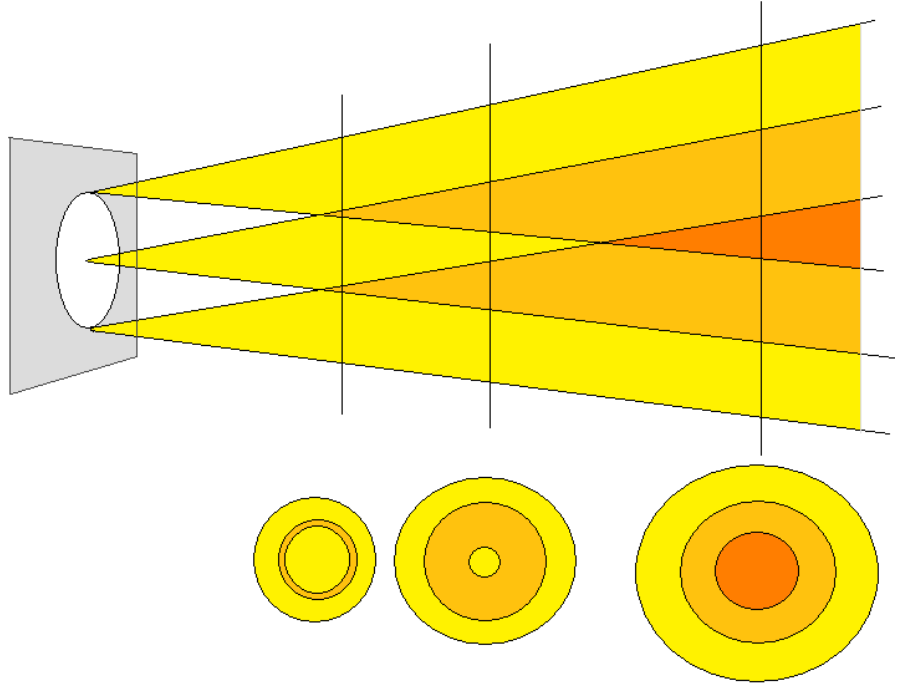,width=12cm}}
\vspace*{0pt}
\caption{The pinhole's image at different focal lenghts, lighter colours correspond to weaker illumination on the projection's screen.}
\end{figure}
Therefore we expect to correct the measured diameter $m$ by a constant amount $r$, a quantity somehow related to the pinhole radius. 
In the dataset of table 1, obatined with a pinhole of 5 mm of diameter, for recovering the expected value for the apparent diameter it is necessary to apply correction factors $r$ to the measure $m$ according to the formula:
\be
D_{true}=m-r.
\ee

\begin{table}
\tbl{Correction factors for the pinhole's images at different focal lengths}
  {\begin{tabular}{rcl}
    \hline
      Measured diamter m [cm]    & Focal lengths f [cm] & Correction factor r [cm] \\ \hline 
3.2 & 296 & 0.415 \\ \hline 
5.3 & 530 & 0.311 \\ \hline
7.3 & 756 & 0.186* \\ \hline
11.8 & 1230 & 0.227 \\ \hline 
    \hline
  \end{tabular}}
\end{table}
The values of the correction factors $r$ are calculated with respect to the expected value for the Sun diameter of 1941 arcsec  for February $20^{th}$, 2000\cite{eff} projected at the various focal lengths utilized.
The value with the star (*) is the minimum only because of the errors of measure of the diameter $m$, repeating that measure other values are to be expected.
The factor $r$ is not constant, and it reaches the minimum value for 12.3 m of focal length. Such value fulfills Rayleigh equation (1). Moreover we can observe with astonishing precision the position of several solar spots, not visible for smaller focal lengths.
\section{Conclusions}
This experiment has been improved either empirically putting all the apparatus groundbased, as theoretically driving the sampling of measures from small to greater focal length in order to diminish the systematic error.
We can conclude that the precision of this method is $\pm ~~30 ~~{\rm arcsec}$, once eliminated the systematic error with the correction factor $r$.
The factor $r$ is to be calculated (or interpolate) for each pinhole diameter and for each focal length, knowing the true value of the Sun diameter (or of another standard source).
The precision $\pm ~~30 ~~{\rm arcsec}$ of this method is confirmed by the successfull observation of the solar spots.
That precision is enough for detecting the Earth orbit ellipticity, and it is the best one in Astronomy without using the telescope.
That procedures followed for the visible light are the same in the case of X-rays.  
Pinhole devices have been used in X-ray astronomy for the imaging in such waveband, and that experience can straighforward introduce such techniques to the students.
\section*{Acknowledgments}
Thanks to prof. Rosa Satullo, to the students of the ITC Darwin in Rome, Italy, and to dr. Renato Klippert.


\begin{thebibliography}{99}
\bibitem{lom} Lombardi, A. M., I grandi della Scienza, n. 13, Le Scienze, Milano, 2000.
\bibitem{eff}Downey, E. C., Ephem - an interactive astronomical ephemeris program, 1992. {\rm http://www.icra.it/solar/ephemvga.zip}
\end{thebibliography}
\end{document}